\documentclass[sort&compress]{aip-cp}

\usepackage{amssymb}
\usepackage{bm}
\usepackage[numbers]{natbib}
\usepackage{graphicx}
\usepackage{color}

\newcommand{\text}[1]{\textrm{\scriptsize{#1}}}
\newcommand{\eqref}[1]{(\ref{#1})}

\newcommand\beq{\begin{equation}}
\newcommand\eeq{\end{equation}}
\newcommand\beqa{\begin{eqnarray}}
\newcommand\eeqa{\end{eqnarray}}
\newcommand{\dd}{\mathrm{d}}

\newcommand{\vv}{\textsl{v}}

\newcommand{\al}{\alpha}

\usepackage{color}

\begin{document}

\title{Mass transport of driven inelastic Maxwell mixtures}


\author[aff1]{Nagi Khalil}
\author[aff2,aff3]{Vicente Garz\'o\corref{cor1}
}
\eaddress[url]{http://www.eweb.unex.es/eweb/fisteor/vicente}
\affil[aff1]{IFISC (CSIC-UIB), Universitat de les Illes Balears, E-07122 Palma de Mallorca, Spain.}
\affil[aff2]{Departamento de F\'isica, Universidad de Extremadura, 06006 Badajoz, Spain.}
\affil[aff3]{Instituto de Computaci\'on Cient\'ifica Avanzada (ICCAEx), Universidad de Extremadura, 06006 Badajoz, Spain.}
\corresp[cor1]{Corresponding author: vicenteg@unex.es}

\maketitle

\begin{abstract}
Mass transport of a driven granular binary mixture is analyzed from the inelastic Boltzmann kinetic equation for inelastic Maxwell models (IMM). The mixture is driven by a thermostat constituted by two terms: a stochastic force and a drag force proportional to the particle velocity. The combined action of both forces attempts to mimic the interaction of solid particles with the interstitial surrounding gas. As with ordinary gases, the use of IMM allows us to exactly evaluate the velocity moments of the Boltzmann collision operator and so, it opens up the possibility of obtaining the exact forms of the Navier--Stokes transport coefficients of the granular mixture. In this work, the diffusion coefficients associated with the mass flux are explicitly determined in terms of the parameters of the mixture. As a first step, the steady homogeneous state reached by the system when the energy lost by collisions is compensated for by the energy injected by the thermostat is addressed. In this steady state, the ratio of kinetic temperatures are determined and compared against molecular dynamics simulations for inelastic hard spheres (IHS). The comparison shows an excellent agreement, even for strong inelasticity and/or disparity in masses and diameters. As a second step, the set of kinetic equations for the mixture is solved by means of the Chapman-Enskog method for states near homogeneous steady states. In the first-order approximation, the mass flux is obtained and the corresponding diffusion transport coefficients identified. The results show that the predictions for IMM obtained in this work coincide with those previously derived for IHS in the first-Sonine approximation when the non-Gaussian corrections to the zeroth-order approximation are neglected.
\end{abstract}


\section{INTRODUCTION}

In the last years, the kinetic theory of molecular gases has been properly adapted to describe granular matter under rapid flow conditions \cite{BP04}. In these conditions, the motion of grains is quite similar to the random motion of atoms or molecules of an ordinary or conventional gas and so, the tools of kinetic theory modified to account for the inelastic character of collisions can be employed to analyze granular flows. Usually, two different new ingredients are introduced in the modified theory. First, given that the collisions are inelastic, the Boltzmann collision operator must be conveniently changed. Second, given that the energy of the gas monotonically decreases in time, the dynamics stops after few collisions per particle unless some external source of energy is introduced to keep the system in rapid flow conditions.

Regarding the above first point, a simple but realistic model for a granular gas is a gas of inelastic hard spheres. In the simplest version, the spheres are assumed to be completely smooth and hence, the inelasticity of collisions is accounted for by a (positive) coefficient of normal restitution that only affects to the translational degrees of freedom of grains. Thus, in this model (inelastic hard spheres, IHS) the particles loose a fraction of their translational kinetic energy after instantaneous collisions. With respect to the above second point, although in the experiments the energy input is done by driving the system through the boundaries \cite{YHCMW02} or alternatively by bulk driving \cite{AD06,SGS05}, a simple way of heating the system is via the introduction of external nonconservative forces (``thermostats'') \cite{EM90}. When the energy lost by collisions is balanced by the energy injected by the thermostat, the system achieves a non-equilibrium steady state. Although several kind of thermostats have been proposed in the literature \cite{NE98,MS00}, we will assume in this work that the granular mixture is driven by the action of a thermostat composed by two different terms: (i) a drag force proportional to the velocity of the particle, and (ii) a stochastic force with the form of a Gaussian white noise where the particles are randomly kicked between collisions \cite{WM96}. In this case, the corresponding kinetic equations for the mixture have the structure of Fokker--Planck equations plus the corresponding Boltzmann collision operators. It is interesting to note that while the stochastic force tries to model the energy transfer from the surrounding gas to the granular particles, the drag force models the friction of grains with the interstitial viscous gas. The generality of this thermostat relays on the fact that it results as the limiting behavior of different ways of kicking the grains \cite{khga14}. In addition, for the sake of simplicity, it is assumed that the collision dynamics of grains is not affected by the thermostat (or equivalently, by the interstitial gas). This requires that the mean-free time between collisions is much less than the time taken by the fluid forces to significantly affect the motion of solid particles.

On the other hand, as in the case of elastic collisions \cite{C88,CC70}, the intricate mathematical structure of the Boltzmann operator for IHS prevents the possibility of obtaining exact results. In particular, the Navier--Stokes transport coefficients of driven granular mixtures have been \emph{approximately} obtained by considering the leading terms in a Sonine polynomial expansion \cite{khga13,khga18}. As Maxwell realized for elastic collisions \cite{C88}, a possible way of overcoming the mathematical difficulties embodied in the hard-sphere kernel is to assume scattering laws where the collision rate of two particles is independent of their relative velocity. This sort of models are referred to as inelastic Maxwell models (IMM) and were introduced in granular literature many years ago \cite{BK00,BCG00}. The IMM share with elastic Maxwell molecules that the collision rate is velocity independent, but their collision rules are the same as for IHS. Although IMM are less realistic than IHS, as for ordinary gases \cite{SG95,GS03}, the use of IMM allows us to make progresses in non-equilibrium problems where the use of IHS prevents the possibility of achieving analytical results. Thus, as an example, recently the complete set of Burnett transport coefficients have been explicitly obtained \cite{khgasa14}.

The aim of this paper is to evaluate the diffusion transport coefficients associated with the mass flux of a binary mixture driven by a stochastic bath with friction. As said before, the Navier--Stokes hydrodynamic equations of a binary mixture driven by this  thermostat have been recently derived \cite{khga13,khga18} by solving the Boltzmann equation for IHS by means of the Chapman--Enskog method \cite{CC70}. The derivation of the hydrodynamic equations with explicit forms for the transport coefficients needs not only the approximations required in the free evolving case \cite{GD02,GMD06}, but also other related to the time dependence of the distribution functions close to the steady states. The latter is far from being a trivial point and is an important aspect to take into account upon obtaining the transport properties of the system.
Here, we revisit the problem by starting from the Boltzmann equation for IMM. Our objective is two-fold. First, we want to get the exact forms of the mass flux without introducing additional and sometimes uncontrolled approximations. Secondly, we want to compare the present results for IMM with those derived before for IHS \cite{khga13,khga18}. This comparison will allow us to gauge the degree of confidence of IMM to unveil in a clean way the impact of inelasticity on granular flows. 

The plan of the paper is as follows. First, we  consider the steady homogeneous states and derive equations for the partial temperatures. The theoretical results are compared against molecular dynamics simulations of IHS \cite{khga14}. Then, the Boltzmann kinetic equation is solved by means of the Chapman--Enskog method up to first order and the expressions of the diffusion transport coefficients are explicitly derived. Finally, we end the paper with a brief discussion of the results.


\section{KINETIC DESCRIPTION \label{sec:2}}

We consider a granular binary mixture modeled as a binary mixture of IMM in $d$ dimensions with masses $m_i$ and diameters $\sigma_i$ $(i=1,2)$. The inelasticity of collisions among all pairs is characterized by three independent (positive) constant coefficients of normal restitution $\al_{11}$, $\al_{22}$, and $\al_{12}=\al_{21}$, where $\al_{ij}\leq 1$. Here, $\al_{ij}$ is the coefficient of normal restitution for collisions between particles of species $i$ and $j$. We also assume that the mixture is fluidized by means of an external force or thermostat composed by two terms: (i) a stochastic force assumed to be the form of a Gaussian white noise and (ii) a drag force proportional to the velocity of the particle. Under these conditions, the nonlinear Boltzmann equation for the one-particle distribution function $f_i(\mathbf{r},\mathbf{v},t)$ of species $i$ having a position $\mathbf r$ and a velocity $\mathbf v$ at time $t$ reads \cite{khga14,khga13}
\beq
\label{1}
\partial_{t}f_i+\mathbf{v}\cdot \nabla f_i-\frac{\gamma_\text{b}}{m_i^{\beta}}\Delta \mathbf{U} \cdot
\frac{\partial f_i}{\partial\mathbf{v}}-\frac{\gamma_\text{b}}{m_i^{\beta}}
\frac{\partial}{\partial\mathbf{v}}\cdot \mathbf{V}
f_i-\frac{1}{2}\frac{\xi_\text{b}^2}{m_i^{\lambda}}\frac{\partial^2 f_i}{\partial \vv^2}=\sum_{j=1}^2\; J_{ij}^{\text{IMM}}[\mathbf{v}|f_i,f_j],
\eeq
where $\gamma_\text{b}$ is the drag or friction constant, $\xi_\text{b}^2$ is related to the strength of the stochastic part of the thermal bath, and $\beta$ and $\lambda$ are arbitrary constants of the driven model. Furthermore, $\Delta \mathbf U= \mathbf U-\mathbf U_g$ is the mean flow velocity of the granular gas $\mathbf U$ with respect to the mean flow velocity of the surrounding gas $\mathbf U_g$ and $\mathbf V= \mathbf v-\mathbf U$ is the peculiar velocity, where
\beq
\label{1.1}
\mathbf{U}=\rho^{-1}\sum_{i=1}^2\int\dd \mathbf{v}\; m_{i}\mathbf{v}f_i(\mathbf{r},\mathbf{v},t).
\eeq
In Eq.\ \eqref{1.1}, $\rho=\sum_i m_i n_i$ is the total mass density and 
\beq
\label{1.2}
n_{i}=\int \dd\mathbf{ v}\;f_i(\mathbf{r},\mathbf{v},t)
\eeq
is the local number density of species $i$. Apart from the fields $n_i$ and $\mathbf{U}$, another important hydrodynamic quantity is the granular temperature $T$ defined as 
\beq
\label{1.3}
T=\frac{1}{n}\sum_{i=1}^2\int \dd\mathbf{v}\;\frac{m_{i}}{d}V^{2}f_i(\mathbf{r},\mathbf{v},t),
\eeq
where $n=n_{1}+n_{2}$ is the total number density. At a kinetic level, it is also convenient to introduce the partial kinetic temperatures $T_i$ for each species defined as
\begin{equation}
\label{1.4}
T_i=\frac{m_{i}}{d n_i}\int\; \dd\mathbf{ v}\;V^{2}f_i(\mathbf{r},\mathbf{v},t).
\end{equation}
The partial temperatures $T_i$ measure the mean kinetic energy of each species. According to Eq.\ \eqref{1.3}, the granular temperature $T$ of the mixture can be also written as $T=\sum x_i T_i$, where $x_i=n_i/n$ is the concentration or mole fraction of species $i$.

The difference between IMM and IHS is in the explicit form of the Boltzmann collision operator. The operator $J_{ij}^{\text{IMM}}[f_i,f_j]$ for IMM is obtained from its corresponding version for IHS \cite{BP04} by replacing the magnitude $|\widehat{\boldsymbol{\sigma}}\cdot {\bf g}_{12}|$ with an average term proportional to the granular temperature $T$ but independent of the relative velocity $\mathbf{g}_{12}=\mathbf{v}_1-\mathbf{v}_2$. Here, $\widehat{\boldsymbol{\sigma}}$ is a unit vector directed along the centers of the two colliding spheres. Therefore, the Boltzmann operator for IMM is \cite{G03}
\beq
J_{ij}^{\text{IMM}}\left[{\bf v}_{1}|f_{i},f_{j}\right] =\frac{\omega_{ij}}{n_j\Omega_d}
\int d{\bf v}_{2}\int d\widehat{\boldsymbol{\sigma }}\nonumber\\
\left[ \alpha_{ij}^{-1}f_i(\mathbf{r},\mathbf{v}_1',t)f_{j}(\mathbf{r},\mathbf{v}_2',t)-f_{i}(\mathbf{r},\mathbf{v}_1,t)f_{j}(\mathbf{r},\mathbf{v}_2,t)\right],
\label{2}
\eeq
where $\omega_{ij}\neq \omega_{ji}$ is an effective collision frequency (to be chosen later) for collisions  of type $i$-$j$ and  $\Omega_d=2\pi^{d/2}/\Gamma(d/2)$ is the total solid angle in $d$ dimensions. In addition, the primes on the velocities denote the initial values $\{{\bf v}_{1}^{\prime},{\bf v}_{2}^{\prime}\}$ that lead to $\{{\bf v}_{1},{\bf v}_{2}\}$ following a binary collision:
\begin{equation}
\label{3}
{\bf v}_{1}^{\prime }={\bf v}_{1}-\mu_{ji}\left( 1+\alpha_{ij}
^{-1}\right)(\widehat{\boldsymbol{\sigma}}\cdot {\bf g}_{12})\widehat{\boldsymbol
{\sigma}},
\quad {\bf v}_{2}^{\prime}={\bf v}_{2}+\mu_{ij}\left(1+\alpha_{ij}^{-1}\right) (\widehat{\boldsymbol{\sigma}}\cdot
{\bf g}_{12})\widehat{\boldsymbol{\sigma}},
\end{equation}
where $\mu_{ij}=m_i/(m_i+m_j)$.

The parameters $\beta$ and $\lambda$ can be interpreted as free parameters of the model. In particular, when $\gamma_\text{b}=\lambda=0$, the thermostat employed here reduces to the stochastic thermostat used in previous works \cite{BT02,DHGD02}. On the other hand, if $\beta=1$ and $\lambda=2$, the model reduces to the Fokker--Planck model for ordinary mixtures \cite{puglisi}. In this context, the model of Eq.\ \eqref{1} can be understood as a generalization of previous driven models.

To completely define the collision operator $J_{ij}^{\text{IMM}}$, one has to chose the collision frequencies $\omega_{ij}$. Usually their $\al$-dependence is taken to optimize the agreement with some property of interest for IHS. Of course, the choice is not unique and may depend on the property of interest. In the case of granular mixtures, one takes $\omega_{ij}$ under the criterion that the cooling rate $\zeta$ of IMM is the same as that of IHS (evaluated by using Maxwellian distributions at the partial temperatures $T_i$). This choice yields the result \cite{G03}
\begin{equation}
\label{4}
\omega_{ij}=\frac{\Omega_d}{\sqrt{\pi}} x_j\left(\frac{\sigma_{ij}}{\sigma_{12}}\right)^{d-1}
\left(\frac{\theta_i+\theta_j}{\theta_i\theta_j}\right)^{1/2}\nu_0,
\end{equation}
where $\sigma_{ij}=(\sigma_i+\sigma_j)/2$, $\theta_i= M_i/\chi_i$, and $M_i= m_i/\overline m$. Here, $\overline m= m_1m_2/(m_1+m_2)$ is the reduced mass and $\chi_i= T_i/T$ is the temperature ratio of species $i$. In addition, in Eq.\ \eqref{4}, $\nu_0$ is an effective collision frequency given by
\begin{equation}
\label{5}
\nu_0=n\sigma_{12}^{d-1}\sqrt{\frac{2T}{\overline{m}}}.
\end{equation}

\subsection{STEADY HOMOGENEOUS STATES}

Before considering inhomogeneous states, it is instructive first to analyze the steady \emph{homogeneous} states. In this situation, the partial densities $n_i$ and the granular temperature $T$ are constant and, with an appropriate selection of the frame of reference, $\mathbf{U}=\mathbf{U}_g=\mathbf{0}$. Under these conditions, Eq.\ \eqref{1} becomes
\beq
\label{5.1}
-\frac{\gamma_\text{b}}{m_i^{\beta}}
\frac{\partial}{\partial\mathbf{v}}\cdot \mathbf{V}
f_i-\frac{1}{2}\frac{\xi_\text{b}^2}{m_i^{\lambda}}\frac{\partial^2 f_i}{\partial \vv^2}=\sum_{j=1}^2\; J_{ij}^{\text{IMM}}[\mathbf{v}|f_i,f_j],
\eeq
The steady state condition for the granular temperature ($\partial_t T=0$) can be easily obtained from Eq.\ \eqref{5.1} by multiplying both sides of this equation by $\vv^2$ and integrating over velocity. The result is 
\beq
\label{5.2}
\Lambda \equiv 2\gamma_\text{b}\sum_{i=1}^2
\frac{x_i \chi_i}{m_i^\beta}-\frac{\xi_\text{b}^2}{p}\sum_{i=1}^2\frac{\rho_i}{m_i^\lambda}+\zeta=0,
\eeq
where $\rho_i=m_i n_i$ is the mass density of species $i$, $p=n T$ is the hydrostatic pressure, and 
\begin{equation}
\zeta=-\frac{1}{d n T}\sum_{i=1}^2\sum_{j=1}^2m_i\int \dd\mathbf{ v}\; V^{2}J_{ij}[\mathbf{ v}
|f_{i},f_{j}]  
\label{5.3}
\end{equation}
is the total ``cooling rate'' due to inelastic collisions among all species. Apart from $T$, the set of equations defining the partial temperatures $T_i$ can be derived from Eq.\ \eqref{5.1}. After some manipulations and in dimensionless form, one gets the result
\beq
\label{6}
T^{*}\left[1-(M_i/2)^{\lambda-1-\beta}T_{i}^*\right]\xi^* =M_i^{\lambda-1}\zeta_{i}^* T_{i}^*, \qquad (i=1,2),
\eeq
where $T^*=T/T_\text{b}$, $T_i^*=T_i/T_\text{b}$, and $\xi^*= \xi_\text{b}^2/(\nu_0T\overline m^{\lambda-1})$. Here,  
\beq
\label{7}
T_\text{b}=\frac{\xi_\text{b}^2}{2\gamma_\text{b} (2\overline{m})^{\lambda-\beta-1}}
\eeq
is a reference or ``bath'' temperature. Its name may be justified since it can be considered as a remnant of the temperature of the interstitial ordinary (elastic) gas. In addition, the (reduced) partial cooling rate (which gives the rate of collisional change of the partial temperature $T_i$) $\zeta_i^*=\zeta_i/\nu_0$ is defined as
\begin{equation}
\label{7.1} \zeta_i^*=\sum_{j=1}^2\zeta_{ij}=-
\frac{m_i}{dn_iT_i}\sum_{j=1}^2\int \dd\mathbf{ v}\; V^{2}J_{ij}[{\bf
v}|f_{i},f_{j}].
\end{equation}
In contrast to the results obtained for IHS \cite{khga13}, the cooling rates $\zeta_i^*$ can be \emph{exactly} obtained for IMM. Their explicit forms for homogeneous states are
\beq
\label{8}
\zeta_{i}^*=\frac{4\pi^{(d-1)/2}}{d\Gamma\left(\frac{d}{2}\right)}
\sum_{j=1}^2
x_{j}\mu_{ji}\left(\frac{\sigma_{ij}}{\sigma_{12}}\right)^{d-1}\left(\frac{\theta_i+\theta_j}
{\theta_i\theta_j}\right)^{1/2}(1+\alpha_{ij})
\left[1-\frac{\mu_{ji}}{2}(1+\alpha_{ij})
\frac{\theta_i+\theta_j}{\theta_j}\right].
\eeq
\begin{figure}[!h]
  \centering
  \includegraphics[angle=-90,width=.325\textwidth]{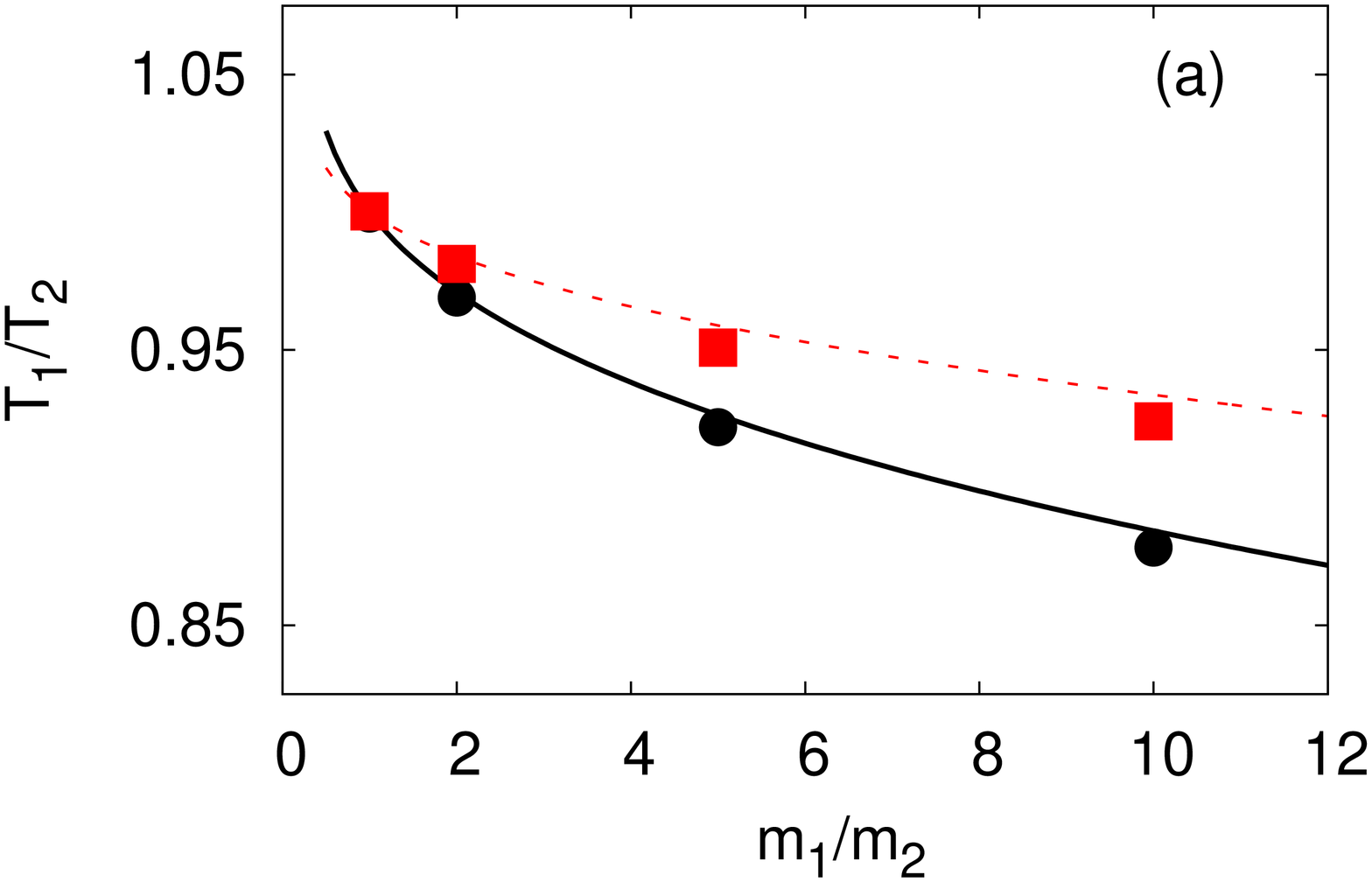}
  \includegraphics[angle=-90,width=.325\textwidth]{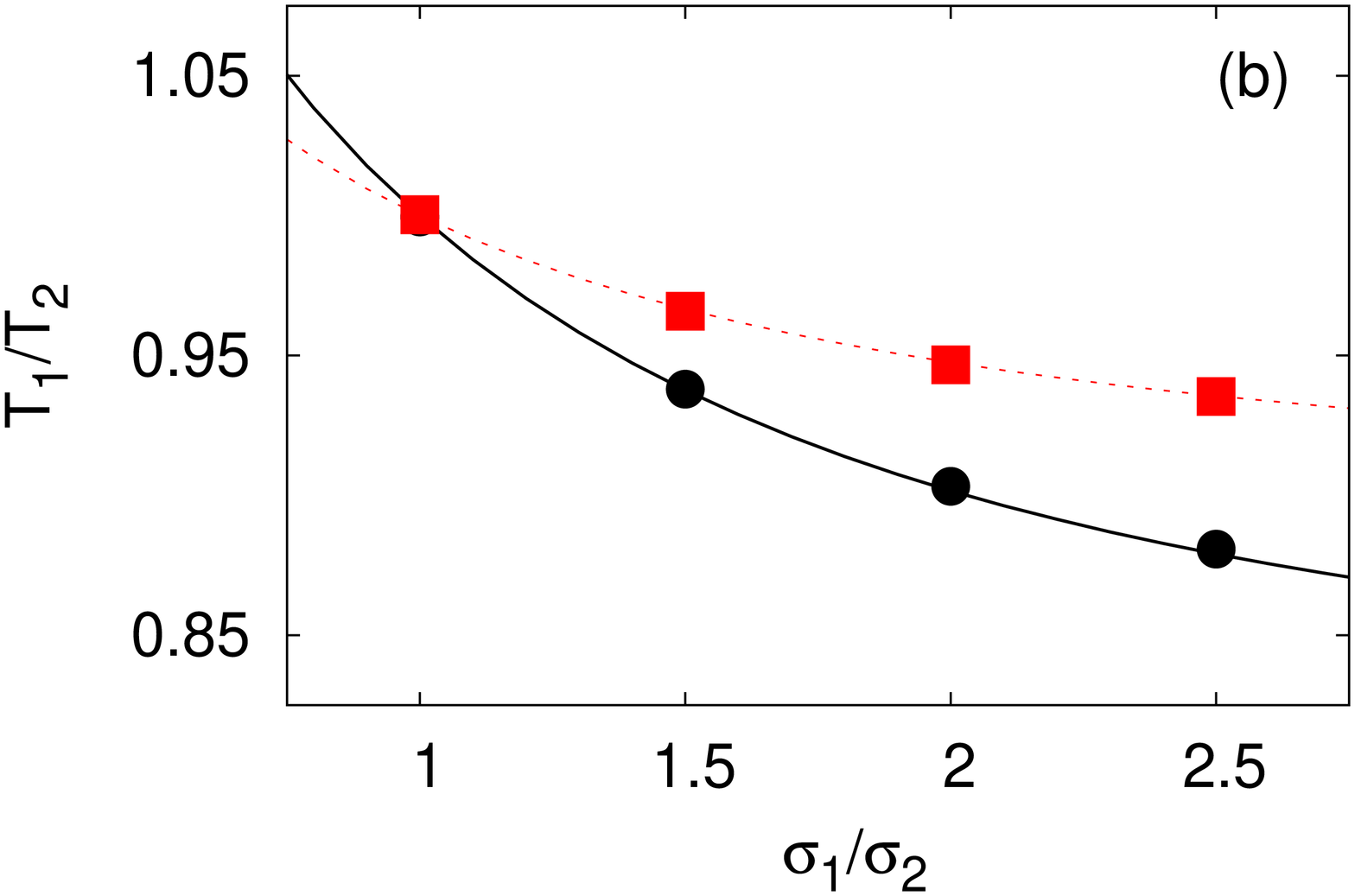}
  \includegraphics[angle=-90,width=.325\textwidth]{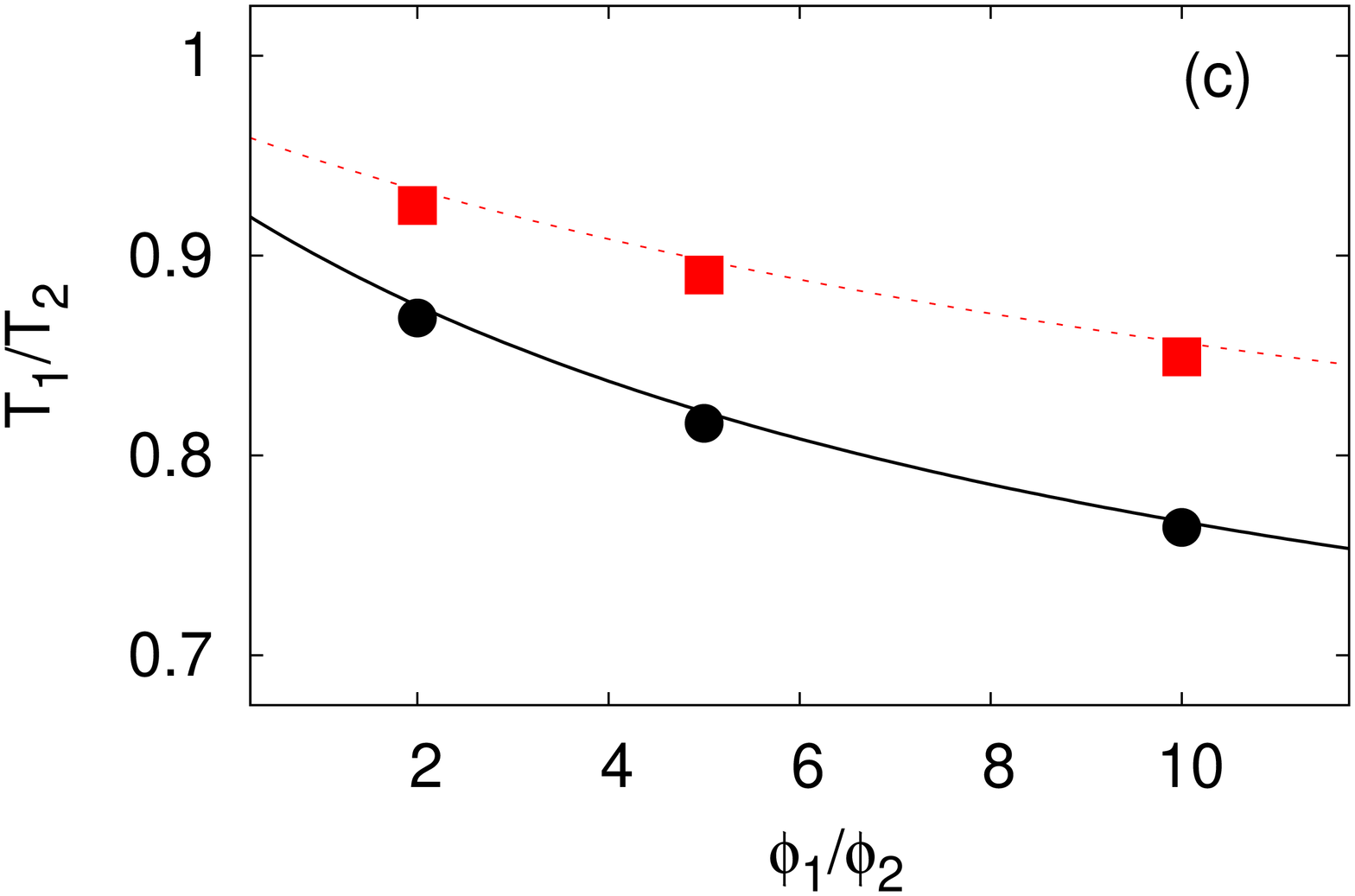}
\caption{Plot of the temperature ratio $T_1/T_2$ as a function of (a) the mass ratio $m_1/m_2$ (for $\sigma_1/\sigma_2=\phi_1/\phi_2 = 1$), (b) the size ratio $\sigma_1/\sigma_2$ (for $m_1/m_2 =\phi_1/\phi_2=1$), and (c) the composition ratio $\phi_1/\phi_2$ (for $m_1/m_2=8$ and $\sigma_1/\sigma_2 = 2$). Here, $d=3$, the volume fraction $\phi = 0.00785$, and two different values of the (common) coefficient of restitution $\al=\al_{11}=\al_{22}=\al_{12}$ have been considered: $\alpha= 0.8$ (solid lines and circles) and $0.9$ (dashed lines and squares). The lines are the theoretical predictions for IMM and the symbols refer to the MD simulation results reported in Ref.\ \cite{khga14}.}
\label{fig:1}
\end{figure}

The partial temperatures $T_i^*$ can be determined by substituting Eq.\ \eqref{8} into the coupled equations \eqref{6}. The solution of this set of equations gives the explicit dependence of the temperature ratio $T_1/T_2$ on the parameter space of the problem (ratios of mass and diameters, concentration, and coefficients of restitution). The temperature ratio provides a measure of the breakdown of the energy equipartition in granular mixtures. It is important to note that the exact results obtained here for the homogeneous steady state for IMM agree with those derived before for IHS \cite{khga13} when non-Gaussian corrections to the distribution functions are neglected. Since the above corrections are in general very small, one expects that the theoretical results for the temperature ratio obtained by solving Eq.\ \eqref{7} present a good agreement with computer simulations. To confirm this expectation, Fig.\ \ref{fig:1} compares the theoretical predictions for IMM with molecular dynamics (MD) simulations of a mixture of inelastic hard spheres \cite{khga14} for a very dilute system (solid volume fraction $\phi=0.00785$). More specifically, we plot $T_1/T_2$ versus the mass ratio $m_1/m_2$ (panel (a)), the size ratio $\sigma_1/\sigma_2$ (panel (b)), and the composition ratio $\phi_1/\phi_2$ (panel (c)), respectively, for two different values of the (common) coefficient of restitution $\al_{ij}=\al$. Here, $\phi_i=n_i \pi \sigma_i^3/6$, $\beta=1$, $\lambda=2$, $\gamma_\text{b}=0.1$, and $\xi_\text{b}^2=0.2$. It is quite apparent that the agreement between theory and simulations is excellent, confirming the reliability of IMM. In addition, we observe that the temperature ratio is a decreasing function of the mass ratio $m_1/m_2$ if the remaining quantities are kept equal for both species. This is not an obvious result since the thermostat differentiates between particles of different masses. A similar behavior is found if the temperature ratio is plotted versus the size ratio $\sigma_1/\sigma_2$ and the composition ratio $\phi_1/\phi_2$.

\section{CHAPMAN--ENSKOG SOLUTION. FIRST-ORDER DISTRIBUTION FUNCTION\label{sec:3}}

Now we consider inhomogeneous situations. More specifically, we consider states that deviate from steady homogeneous states by \emph{small} spatial gradients. In this state, the Boltzmann equation \eqref{1} may be solved by means of the Chapman--Enskog method \cite{CC70} conveniently adapted to account for dissipative dynamics. As usual, the method assumes that, after a transient regime (of the order of a few collision times), the system achieves a hydrodynamic regime where the one-particle distribution functions of each species adopt a \emph{normal} form. This means that all space an time dependence of $f_i(\mathbf{r},\mathbf{v},t)$ only occurs through the hydrodynamic fields. As pointed out in previous works \cite{GD02,GMD06}, there is more flexibility in the choice of the fields in granular mixtures than in the case of ordinary mixtures. Here, as for undriven granular mixtures \cite{GD02,GMD06}, we take the concentration $x_1$, the hydrostatic pressure $p$, the granular temperature $T$, and the $d$ components of the mean flow velocity $\mathbf{U}$ as the $d+3$ independent fields of the binary mixture. Thus, for times longer than the mean free time, the distributions $f_i$ can be written as
\beq
\label{8.1}
f_i(\mathbf{r}, \mathbf{v},t)=f_i\left[\mathbf{v}|x_1(t), p(t), T(t), \mathbf{U}(t)\right].
\eeq
The notation on the right hand side indicates a functional dependence on concentration, pressure, temperature, and flow velocity. In the case of small spatial variations, the functional dependence \eqref{8.1} can be made local in space and time through an expansion in gradients of the fields: 
\begin{equation}
f_{i}=f_{i}^{(0)}+\epsilon \,f_{i}^{(1)}+\epsilon^2 \,f_{i}^{(2)}+\cdots,
\label{8.2}
\end{equation}
where each factor of $\epsilon$ means an implicit gradient of a hydrodynamic field. In addition, in ordering the different level of approximations in the kinetic equations, one has to characterize the magnitude of the driven parameters $\gamma_\text{b}$ and $\xi_\text{b}^2$ relative to the gradients as well. As in the study of IHS \cite{khga13}, they are taken to be of zeroth order in the gradients. This assumption is based on the fact that both driven parameters do not induce any flux in the system. On the other hand, a different consideration must be given to the term proportional to the velocity difference $\Delta \mathbf{U}$. Since this term is expected to contribute to the mass flux (for instance, in sedimentation problems), then $\Delta \mathbf{U}$ must be considered at least to be of first order in perturbation expansion.

To first-order in spatial gradients, the calculations follow similar steps to those carried out before for IHS \cite{khga13}. In particular, the distribution $f_1^{(1)}$ obeys the kinetic equation 
\beqa
\label{9}
&&\partial_{t}^{(0)}f_{1}^{(1)}-\frac{\gamma_\text{b}}{m_1^\beta}
\frac{\partial}{\partial\mathbf{ v}}\cdot \mathbf{V}
f_1^{(1)}-\frac{1}{2}\frac{\xi_\text{b}^2}{m_1^\lambda}\frac{\partial^2}{\partial \vv^2}f_1^{(1)}
+{\cal L}_{1} f_{1}^{(1)}+{\cal M}_{1}f_{2}^{(1)}=\mathbf{ A}_{1}\cdot \nabla x_{1}+\mathbf{ B}_{1}\cdot \nabla p+\mathbf{ C}_{1}\cdot \nabla T\nonumber\\
& &
+D_{1,k\ell}\frac{1}{2}\left(\nabla_{k}U_{\ell}+\nabla_{\ell}U_{k}
-\frac{2}{d}\delta_{k\ell}\nabla \cdot\mathbf{U}\right)+E_1 \nabla \cdot \mathbf{U}+\mathbf{G}_1\cdot \Delta \mathbf{U}.
\eeqa
The coefficients of the field gradients on the right side are functions of $\mathbf{V}$ and the hydrodynamic fields. They are given by
\begin{equation}
\mathbf{A}_{1}(\mathbf{V})=-\mathbf{V}\frac{\partial f_1^{(0)}}{\partial x_1}+\frac{\gamma_\text{b} (m_2^\beta-m_1^\beta)}{\rho^2 (m_1 m_2)^{\beta-1}}\frac{p}{T} D \frac{\partial f_1^{(0)}}{\partial \mathbf{V}},
\label{10}
\end{equation}
\beq
\label{11}
\mathbf{B}_{1}(\mathbf{V})=-\mathbf{V}\frac{\partial f_1^{(0)}}{\partial p} -\rho^{-1}
\frac{\partial f_1^{(0)}}{\partial \mathbf{V}}
+\frac{\gamma_\text{b}(m_2^\beta-m_1^\beta)}{p(m_1 m_2)^\beta} D_p \frac{\partial f_1^{(0)}}{\partial \mathbf{V}},
\eeq
\begin{equation}
\mathbf{C}_{1}(\mathbf{V})=-\mathbf{V} \frac{\partial f_1^{(0)}}{\partial T}+\frac{\gamma_\text{b}(m_2^\beta-m_1^\beta)}{T(m_1m_2)^\beta}D_T \frac{\partial f_1^{(0)}}{\partial \mathbf{V}},
\label{12}
\end{equation}
\begin{equation}
D_{1,k\ell}(\mathbf{V})=V_k \frac{\partial f_1^{(0)} }{\partial V_\ell},
\label{13}
\end{equation}
\beq
\label{14}
E_1(\mathbf{V})=\frac{d+2}{d}p \frac{\partial f_1^{(0)}}{\partial p}+ \frac{2}{d}T\frac{\partial f_1^{(0)}}{\partial T}+\frac{1}{d}\mathbf{V}\cdot \frac{\partial f_1^{(0)}}{\partial \mathbf{V}},
\eeq
\begin{equation}
\label{15}
\mathbf{G}_1(\mathbf{V})=\frac{\gamma_\text{b}}{\rho} \frac{m_2^\beta-m_1^\beta}{(m_1m_2)^\beta}\left(\rho_2+ D_U\right) \frac{\partial f_1^{(0)}}{\partial \mathbf{V}}.
\end{equation}
Moreover,  the linear operators ${\cal L}_1$ and ${\cal M}_{1}$ are defined as  
\begin{equation}
{\cal L}_{1}X=-\left(
J_{11}^{\text{IMM}}[f_{1}^{(0)},X]+J_{11}^{\text{IMM}}[X,f_{1}^{(0)}]+
J_{12}^{\text{IMM}}[X,f_{2}^{(0)}]\right) \;,
\label{15.1}
\end{equation}
\begin{equation}
{\cal M}_{1}X=-J_{12}^{\text{IMM}}[f_{1}^{(0)},X].  
\label{15.2}
\end{equation}
The corresponding equation for $f_2^{(1)}$ is obtained from Eq.\ \eqref{9} by setting $1\leftrightarrow 2$ (except for the field $x_1$). In addition, upon deriving Eqs.\ \eqref{9}--\eqref{15}, use has been made of the form of the mass flux $\mathbf{j}_1^{(1)}$ defined as
\beq
\label{16}
\mathbf{j}_1^{(1)}=\int\; \dd\mathbf{v}\; m_1 \mathbf{V} f_1^{(1)}(\mathbf{V}).
\eeq  
To first-order in spatial gradients (Navier--Stokes hydrodynamic order), the constitutive equation of $\mathbf{j}_1^{(1)}$ is   
\beq
\label{20}
\mathbf{j}_{1}^{(1)}=-\left(\frac{m_{1}m_{2}n}{\rho }\right) D\nabla x_{1}-
\frac{\rho}{p}D_{p}\nabla p-\frac{\rho}{T}D_T\nabla T-D_U \Delta \mathbf{U}, \quad \mathbf{j}_2^{(1)}=-\mathbf{j}_1^{(1)}.
\eeq
Here, $D$ is the diffusion coefficient, $D_p$ is the pressure diffusion coefficient, $D_T$ is the thermal diffusion coefficient, and $D_U$ is the velocity diffusion coefficient. Note that Eq.\ \eqref{9} has the same structure as for IHS \cite{khga13} except for the form of the linearized Boltzmann collision operators $\mathcal{L}_1$ and $\mathcal{M}_1$ and the fact that the first-order contribution to the cooling rate vanishes for IMM (i.e., $\zeta_U=0$ where $\zeta^{(1)}=\zeta_U\nabla \cdot \mathbf U$).

\section{MASS FLUX}
\label{sec:4}

We compute here the first-order contribution $\mathbf{j}_1^{(1)}$ to the mass flux. To get it, we multiply both sides of Eq.\ \eqref{9} by $m_1 \mathbf{V}$ and integrate over velocity. After some algebra, one achieves the result
\beqa
\label{17}
\partial_t^{(0)}\mathbf{j}_1^{(1)}+\frac{\gamma_\text{b}}{m_1^\beta}\mathbf{j}_1^{(1)}+\nu_D \mathbf{j}_1^{(1)}&=&
-\left[p\frac{\partial}{\partial x_1}\left(x_1 \chi_1\right)+\frac{\gamma_\text{b} \rho_1(m_2^\beta-m_1^\beta)}{\rho^2 (m_1 m_2)^{\beta-1}}\frac{p}{T} D \right]\nabla x_1 \nonumber\\
& & -\left[x_1 \left(\chi_1+p\frac{\partial \chi_1}{\partial p}\right)-\frac{\rho_1}{\rho}+\frac{\gamma_\text{b} \rho_1(m_2^\beta-m_1^\beta)}{p (m_1 m_2)^{\beta}}D_p \right]\nabla p \nonumber\\
& & -\left[px_1\frac{\partial \chi_1}{\partial T}+\frac{\gamma_\text{b} \rho_1(m_2^\beta-m_1^\beta)}{T (m_1 m_2)^{\beta}}D_T\right] \nabla T \nonumber\\
& &
-\frac{\gamma_\text{b}\rho_1}{\rho} \frac{m_2^\beta-m_1^\beta}{(m_1m_2)^\beta}\left(\rho_2+ D_U\right)\Delta \mathbf{U}.
\eeqa
Upon obtaining Eq.\ \eqref{17}, use has been made of the result
\beq
\label{18}
\int\; \dd\mathbf{v}\; m_1 \mathbf{V}\left(\mathcal{L}_1f_1^{(1)}+\mathcal{M}_1f_2^{(1)}\right)=\nu_D \mathbf{j}_1^{(1)},
\eeq
where
\beq
\label{19}
\nu_D=\rho \frac{\omega_{12}}{dn_2}\frac{1+\al_{12}}{m_1+m_2}=
\frac{2\pi^{(d-1)/2}}{d\Gamma\left(\frac{d}{2}\right)}(1+\al_{12})\left(\frac{M_1\chi_2+M_2\chi_1}{M_1 M_2}\right)^{1/2}
\left(x_2 M_1^{-1}+x_1 M_2^{-1}\right)\nu_0.
\eeq

The set of diffusion transport coefficients $\left\{D, D_p, D_T, D_U\right\}$ can be obtained by substituting the constitutive equation \eqref{20} into Eq.\ \eqref{17}. Although we are interested in computing the transport coefficients under steady conditions, to consistently retain all the contributions to the mass transport, one assumes first that $\partial_t^{(0)}T$ and $\partial_t^{(0)}p$ are different from zero and then one takes the steady state condition $\Lambda=0$. In order to compute the time derivative $\partial_t^{(0)}\mathbf{j}_1^{(1)}$ in Eq.\ \eqref{17}, dimensional analysis shows that $D\propto T^{1/2}$, $D_p\propto D_T\propto T^{3/2}/p$, and $D_U \propto p/T$. Hence, the coefficient $D_U$ does not depend on time and verifies an autonomous equation. Its explicit form can be easily identified from Eq.\ \eqref{17} as
\beq
\label{30}
D_U=\left[\frac{\gamma_\text{b}}{m_1^\beta}\left(1-\frac{\rho_1}{\rho}\frac{m_2^\beta-m_1^\beta}{m_2^\beta}\right)+\nu_D\right]^{-1}
\frac{\gamma_\text{b}\rho_1\rho_2}{\rho}\frac{m_2^\beta-m_1^\beta}{(m_1m_2)^\beta}.
\eeq
When $m_1=m_2$ or $\gamma_b= 0$, then $D_U= 0$ as expected from the previous result obtained for IHS \cite{khga13}.

On the other hand, the remaining three coefficients $D$, $D_p$, and $D_T$ obey a set of coupled linear algebraic equations in the steady state. To obtain them, one has to take into account the intermediate results
\beq
\label{22}
\partial_t^{(0)}\nabla p=\nabla(\partial_t^{(0)} p)=-\nabla(p\Lambda)=-\Lambda\nabla p+\mathcal{P}_{x_1}\nabla x_1+\mathcal{P}_p \nabla p+\mathcal{P}_T \nabla T,
\eeq
\beq
\partial_t^{(0)}\nabla T=\nabla(\partial_t^{(0)} T)=-\nabla(T\Lambda)=-\Lambda \nabla T+\mathcal{T}_{x_1}\nabla x_1+\mathcal{T}_p \nabla p+\mathcal{T}_T \nabla T,
\eeq
where $\Lambda=\partial_t^{(0)} \ln T=\partial_t^{(0)} \ln p$ is defined by Eq.\ \eqref{5.2}. Here, we have introduced the following auxiliary functions:
\beq
\label{23}
\mathcal{P}_{x_1}=\frac{p}{T}\xi_\text{b}^2 \frac{m_2^{\lambda-1}-m_1^{\lambda-1}}{(m_1m_2)^{\lambda-1}}-p \frac{\partial \zeta}{\partial x_1}-2\gamma_\text{b} p\frac{m_2^\beta-m_1^\beta}{(m_1m_2)^\beta}\left(\chi_1+x_1\frac{\partial \chi_1}{\partial x_1} \right),
\eeq
\beq
\label{24}
\mathcal{P}_p=-\left(2\gamma_\text{b}\sum_{i=1}^2 \frac{x_i\chi_i}{m_i^\beta}+2\gamma_\text{b} p\frac{m_2^\beta-m_1^\beta}{(m_1m_2)^\beta}x_1 \frac{\partial \chi_1}{\partial p}-\xi_\text{b}^2\frac{1}{T}\sum_{i=1}^2\frac{x_i}{m_i^{\lambda-1}} +\zeta+p\frac{\partial \zeta}{\partial p}\right),
\eeq
\beq
\label{25}
\mathcal{P}_T=-\left(\xi_\text{b}^2\frac{p}{T^2}\sum_{i=1}^2\frac{x_i}{m_i^{\lambda-1}}+p\frac{\partial \zeta}{\partial T} +2\gamma_\text{b} p\frac{m_2^\beta-m_1^\beta}{(m_1m_2)^\beta}x_1 \frac{\partial \chi_1}{\partial T}
\right),
\eeq
\beq
\label{26}
\mathcal{T}_{x_1}=\xi_\text{b}^2 \frac{m_2^{\lambda-1}-m_1^{\lambda-1}}{(m_1m_2)^{\lambda-1}}-T \frac{\partial \zeta}{\partial x_1}-2\gamma_\text{b} T\frac{m_2^\beta-m_1^\beta}{(m_1m_2)^\beta}\left(\chi_1+x_1\frac{\partial \chi_1}{\partial x_1} \right),
\eeq
\beq
\label{27}
\mathcal{T}_p=-\left(2\gamma_\text{b} T\frac{m_2^\beta-m_1^\beta}{(m_1m_2)^\beta}x_1 \frac{\partial \chi_1}{\partial p}+T\frac{\partial \zeta}{\partial p}\right),
\eeq
\beq
\label{28}
\mathcal{T}_T=-\left(2\gamma_\text{b} \sum_{i=1}^{2}\frac{x_i\chi_i}{m_i^\beta}+2\gamma_\text{b} T\frac{m_2^\beta-m_1^\beta}{(m_1m_2)^\beta}x_1 \frac{\partial \chi_1}{\partial T}+\zeta+T\frac{\partial \zeta}{\partial T}\right).
\eeq
Therefore, in the steady state ($\Lambda=0$), the zeroth-order time derivative $\partial_t^{(0)}\mathbf{j}_1^{(1)}$ gives the contributions
\beq
\label{29}
\partial_t^{(0)}\mathbf{j}_1^{(1)} \to -\left(\frac{\rho}{p}\mathcal{P}_{x_1}D_p+\frac{\rho}{T}\mathcal{T}_{x_1}D_T\right)\nabla x_1
-\left(\frac{\rho}{p}\mathcal{P}_{p}D_p+\frac{\rho}{T}\mathcal{T}_{p}D_T\right)\nabla p
-\left(\frac{\rho}{p}\mathcal{P}_{T}D_p+\frac{\rho}{T}\mathcal{T}_{T}D_T\right)\nabla T.
\eeq

Since the gradients of the hydrodynamic fields are all independent, the set of equations defining $D$, $D_p$, and $D_T$ can be easily obtained when one substitutes Eqs.\ \eqref{20} and \eqref{29} into Eq.\ \eqref{17} and separates the terms corresponding to the spatial gradients $\nabla x_1$, $\nabla p$, and $\nabla T$. After simple algebra, one gets the following set of \emph{coupled} linear algebraic equations:
\beq
\label{31}
\left[\frac{m_1m_2n}{\rho}\left(\frac{\gamma_\text{b}}{m_1^\beta}+\nu_D\right)-\frac{\gamma_\text{b} \rho_1(m_2^\beta-m_1^\beta)}{\rho^2 (m_1 m_2)^{\beta-1}}\frac{p}{T}\right]D+
\frac{\rho}{p}\mathcal{P}_{x_1}D_p+\frac{\rho}{T}\mathcal{T}_{x_1}D_T=p\frac{\partial}{\partial x_1}\left(x_1 \chi_1\right),
\eeq
\beq
\label{32}
\left[\frac{\rho}{p}\mathcal{P}_{p}+\frac{\rho}{p}\left(
\frac{\gamma_\text{b}}{m_1^\beta}+\nu_D\right)-\frac{\gamma_\text{b} \rho_1(m_2^\beta-m_1^\beta)}{p (m_1 m_2)^{\beta-1}}\right]D_p
+\frac{\rho}{T}\mathcal{T}_{p}D_T=x_1 \chi_1\left(1-\frac{\rho_1}{\rho x_1\chi_1}\right)+x_1p\frac{\partial \chi_1}{\partial p},
\eeq
\beq
\label{33}
\frac{\rho}{p}\mathcal{P}_{T}D_p+\left[\frac{\rho}{T}\mathcal{T}_{T}+\frac{\rho}{T}\left(
\frac{\gamma_\text{b}}{m_1^\beta}+\nu_D\right)-\frac{\gamma_\text{b} \rho_1(m_2^\beta-m_1^\beta)}{T (m_1 m_2)^{\beta-1}}\right]
D_T=x_1p\frac{\partial \chi_1}{\partial T}.
\eeq
The solution to Eqs.\ \eqref{31}--\eqref{33} gives $D$, $D_p$, and $D_T$ in terms of the parameter space of the problem. In addition, the explicit expressions of the above coefficients depend on the derivatives of the cooling rate $\partial \zeta/\partial x_1$, $\partial \zeta/\partial p$, and $\partial \zeta/\partial T$ as well as the derivatives of the temperature ratio $\partial \chi/\partial x_1$, $\partial \chi/\partial p$, and $\partial \chi/\partial T$. The dependence of these derivatives on the parameter space can be found in Ref.\ \cite{khga13}.  Given that the explicit forms of the above transport coefficients are very long, it will be omitted here for the sake of brevity.
\begin{figure}[!h]
  \centering
  \includegraphics[angle=-90,width=.325\textwidth]{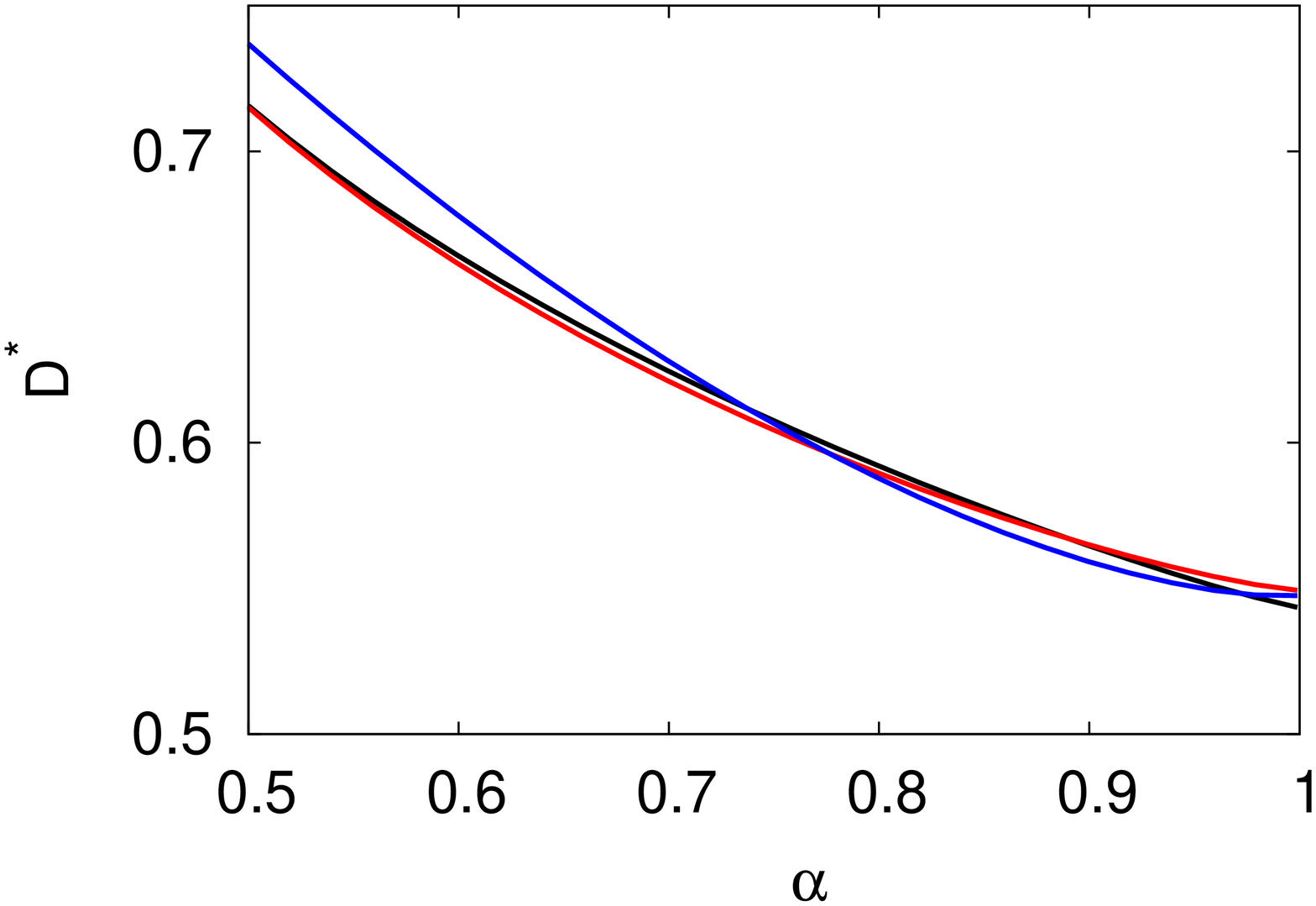}
  \includegraphics[angle=-90,width=.325\textwidth]{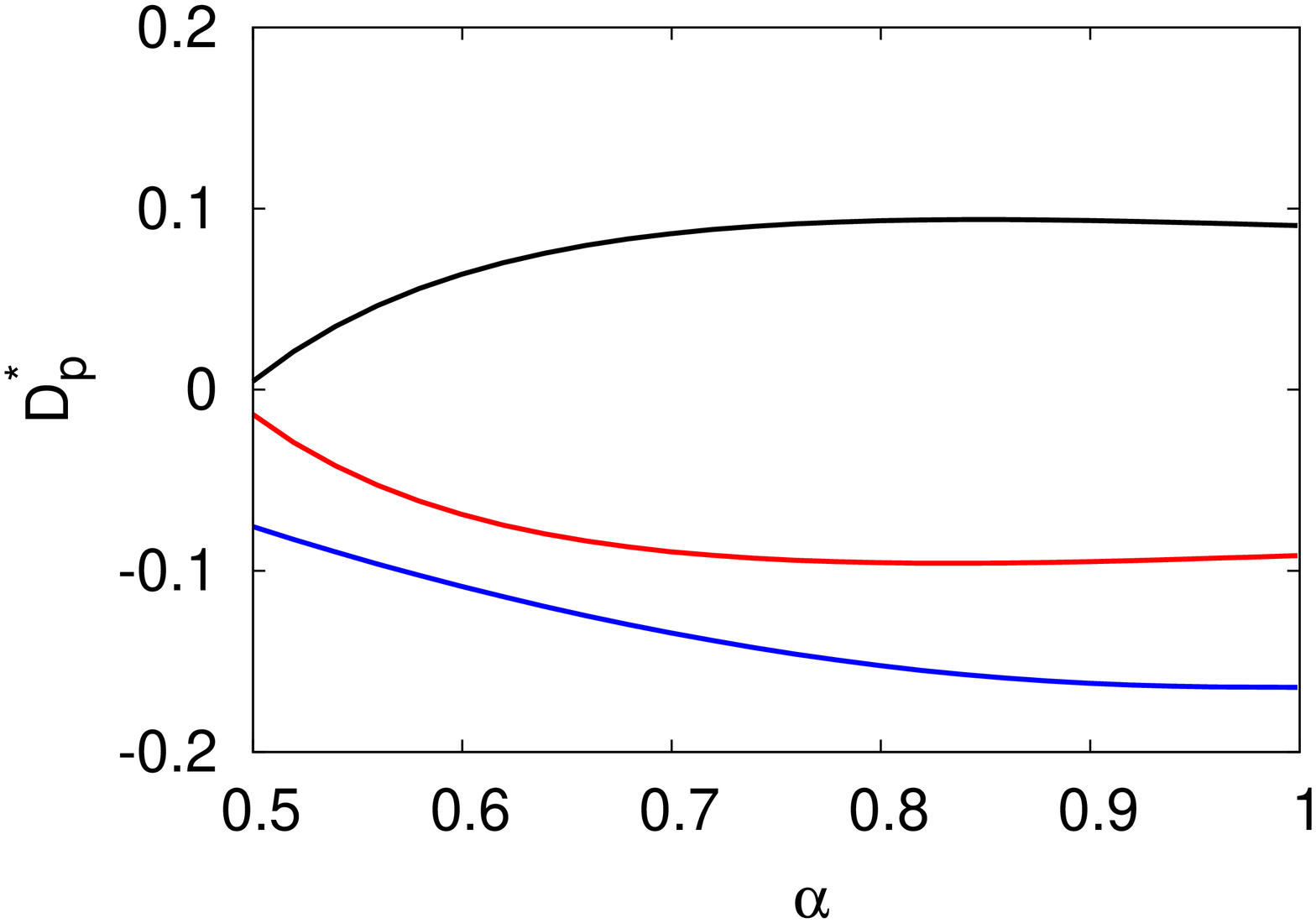}
  \includegraphics[angle=-90,width=.325\textwidth]{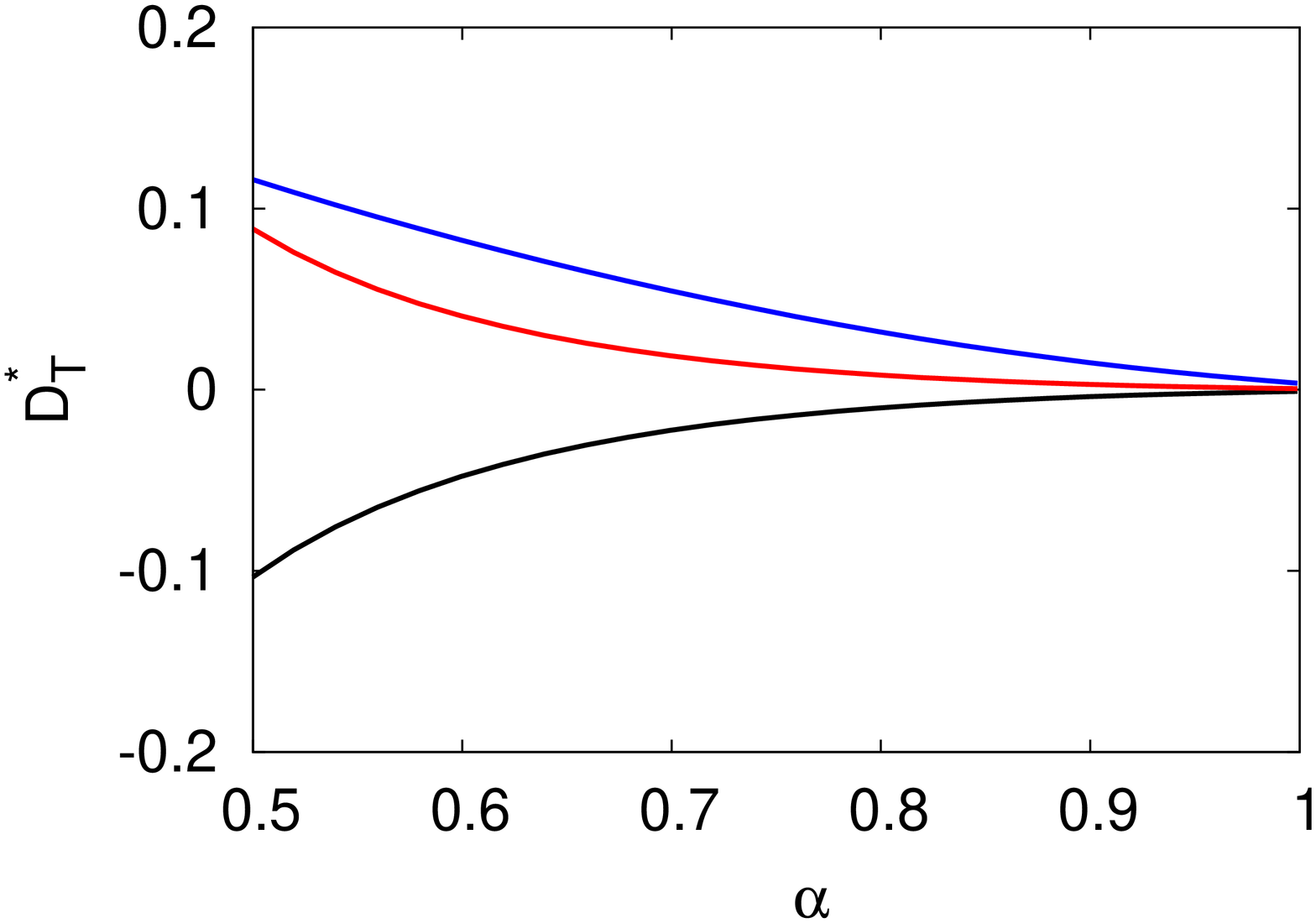}
\caption{Reduced diffusion coefficients $D^*$, $D_p^*$, and $D_T^*$ as a function of the (common) coefficient of restitution $\alpha_{11} =\alpha_{12} =\alpha_{22}=\alpha$ for an equimolar binary mixture ($x_1=\frac{1}{2}$) of hard disks ($d=2$) with $\sigma_1/\sigma_2=1$, and $m_1/m_2=0.5$ (black), $2$ (red), $4$ (blue). The parameters of the driven model are $\xi_b^2=0.2$, $\gamma_b=0.1$, $\lambda=2$, and $\beta=1$.}
\label{fig2}
\end{figure}

As for the temperature ratio, the results of IMM for $D$, $D_p$, $D_T$, and $D_U$ coincide with those previously derived for IHS in the first Sonine approximation \cite{khga13}. This shows again the degree of reliability of IMM to gauge the impact of inelasticity on mass transport. To illustrate the dependence of the diffusion transport coefficients on the coefficients of restitution, it is convenient to introduce the dimensionless forms
\begin{equation}
D^*=\frac{m_{1}m_{2}\nu_{0}}{\rho T}D,\quad
D_{p}^*=\frac{\rho\nu_{0}}{nT}D_{p},\quad
D_T^*=\frac{\rho \nu_{0}}{nT}D_T.
\end{equation}
Figure \ref{fig2} shows $D$, $D_p^*$, and $D_T^*$ versus the (common) coefficient of restitution $\al$ for a two-dimensional mixture with $x_1=1/2$, $\sigma_1=\sigma_2$, and three different values of the mass ratio $m_1/m_2$. As expected, we observe first that the role played by the thermostat is not neutral since the $\al$-dependence of the transport diffusion coefficients is different from the one found for undriven granular mixtures \cite{GMD06}. In particular, while the (reduced) pressure diffusion coefficient $D_p^*$ exhibits a monotonic increase with decreasing $\al$ in all cases for undriven mixtures (see Fig.\ 2 of Ref.\ \cite{GMD06}), Fig.\ \ref{fig2} shows that $D_p^*$ increases (decreases) with inelasticity when $m_1>m_2$ ($m_1<m_2$) for driven mixtures. The opposite dependence on the mass ratio is found for the thermal diffusion coefficient $D_T^*$ in the case of undriven mixtures. Regarding the diffusion transport coefficient $D^*$, we see that this coefficient is a decreasing function of the coefficient of restitution with a weak dependence on the mass ratio. In addition, Fig.\ \ref{fig2} also highlights that the impact of inelasticity on mass transport is fewer than the one reported before for undriven granular mixtures \cite{GMD06}.

\section{DISCUSSION}
\label{sec:5}

In this paper, we have addressed the determination of the diffusion transport coefficients of a granular binary mixture driven by a stochastic bath with friction. These coefficients have been obtained by solving the Boltzmann kinetic equation for IMM via the Chapman--Enskog method adapted to dissipative dynamics. In contrast to previous attempts carried out for IHS \cite{khga13,khga14,khga18}, the theoretical results are exact since the collisional moments needed to get the transport coefficients have been exactly computed without the knowledge of the distribution functions.

Before considering transport properties, the \emph{steady} homogeneous states have been studied. In this state, the temperature ratio $T_1/T_2$ (which measures the departure of energy equipartition) has been obtained in terms of the driven parameters, the mechanical parameters of the mixture (masses, diameters, and coefficients of restitution), and the concentration. A comparison between the theoretical predictions of $T_1/T_2$ of IMM with previous MD simulations of IHS \cite{khga14} show an excellent agreement, even for conditions of strong inelasticity and/or disparate masses. Moreover, the analytical results for IMM coincide with those derived for IHS \cite{khga13} when non-Gaussian corrections to the distribution functions of each species are neglected.

Once the steady homogeneous states are well characterized, the Boltzmann equation is solved by means the Chapman--Enskog expansion around the \emph{local} version of the homogeneous solution (zeroth-order approximation). As already noted in the previous works devoted to IHS \cite{khga13,khga18}, a subtle point in the derivation is that the presence of an external thermostat introduces the possibility of a local energy unbalance. This unbalance gives rise to new contributions to the transport coefficients which were not accounted for in previous studies \cite{GM02} where a local steady state was assumed for the zeroth-order distribution. In this paper, we have focused on mass transport where four relevant transport coefficients (diffusion, pressure diffusion, thermal diffusion, and velocity diffusion) have been identified. As in the case of the temperature ratio, these transport coefficients have been exactly determined in terms of the parameter space of the system. An interesting result is that the expressions of the diffusion transport coefficients of IMM are the same as those derived before for IHS in the first Sonine approximation \cite{khga13}. Therefore, one could conclude that the excellent agreement found here for driven mixtures between IMM and IHS (at the level of temperature ratio and mass transport) can be considered again as a testimony of the confidence of the first model to display the role played by inelasticity in granular mixtures. An interesting future work is to determine the remaining transport coefficients of the mixture (the shear viscosity and those associated with the heat flux) to measure the accuracy of IMM in higher order moments. Previous results \cite{GA05} derived for undriven mixtures have shown discrepancies between both interaction models for the above moments.


\section{ACKNOWLEDGMENTS}
The research of N.K. and V.G. has been supported by the Spanish Agencia Estatal de Investigaci\'on through Grants No. FIS2015-63628-C2-2-R and No. FIS2016-76359- P, respectively, both partially financed by ``Fondo Europeo de Desarrollo Regional'' funds. The research of V.G. has also been supported by the Junta de Extremadura (Spain) through Grant No. GR18079, partially financed by ``Fondo Europeo de Desarrollo Regional'' funds.



\bibliographystyle{aipproc}   


\begin{thebibliography}{11}

\bibitem{BP04}N.~Brilliantov and T. P\"oschel, \emph{Kinetic Theory of Granular Gases} (Oxford University Press, Oxford, 2004).

\bibitem{YHCMW02}X. Yang, C. Huan, D. Candela, R. W. Mair and R. L. Walsworth, Phys. Rev. Lett. \textbf{88}, 044301 (2002);
C. Huan, X. Yang, D. Candela, R. W. Mair, and R. L. Walsworth, Phys. Rev. E \textbf{69}, 041302 (2004).

\bibitem{AD06}A. R. Abate and D. J. Durian, Phys. Rev. E \textbf{74}, 031308 (2006).

\bibitem{SGS05}M. Schr\"oter, D. I. Goldman, and H. L. Swinney, Phys. Rev. E \textbf{71}, 030301(R) (2005).

\bibitem{EM90}D. J. Evans and G. P. Morriss, \emph{Statistical Mechanics of Nonequilibrium Liquids} (Aademic Press, London, 1990).

\bibitem{NE98}T. P. C. van Noije, and M. H. Ernst, Granular Matter \textbf{1}, 57--64 (1998).

\bibitem{MS00}J.M. Montanero and A. Santos, Granular Matter \textbf{2}, 53--64 (2000).

\bibitem{WM96}D. R. M. Williams and F. C. MacKintosh, Phys. Rev. E {\bf 54}, R9--R12 (1996).

\bibitem{khga14}N.~Khalil, and V.~Garz\'o, J. Chem. Phys. \textbf{140}, 164901 (2014).

\bibitem{C88}C. Cercignani, \emph{The Boltzmann Equation and its Applications} (Springer--Verlag, New York, 1988).

\bibitem{CC70}S. Chapman and T. G. Cowling, {\em The Mathematical Theory of Nonuniform Gases} (Cambridge University Press, Cambridge, 1970).

\bibitem{khga13}N.~Khalil, and V.~Garz\'o, Phys. Rev. E \textbf{88}, 052201 (2013).

\bibitem{khga18}N.~Khalil, and V.~Garz\'o, Phys. Rev. E \textbf{97}, 022902 (2018).


\bibitem{BK00}E. Ben-Naim and P. L. Krapivsky, Phys. Rev. E \textbf{61}, R5--R8 (2000).

\bibitem{BCG00}A. V. Bobylev, J. A. Carrillo, and I. Gamba, J. Stat. Phys. \textbf{98} 743--773 (2000).

\bibitem{SG95}A. Santos and V. Garz\'o, Physica A \textbf{213}, 409--425 (1995).



\bibitem{GS03}V. Garz\'o and A. Santos, {\em Kinetic Theory of Gases in Shear Flows. Nonlinear Transport} (Kluwer Academic, Dordrecht, 2003).


\bibitem{khgasa14}N.~Khalil, and V.~Garz\'o, A.~Santos, Phys. Rev. E \textbf{89}, 052201 (2014).

\bibitem{GD02}V. Garz\'o and J. W. Dufty, Phys. Fluids {\bf 14}, 1476--1490 (2002).

\bibitem{GMD06} V. Garz\'o, J. M. Montanero, and J. W. Dufty, Phys. Fluids {\bf 18}, 083305 (2006).

\bibitem{G03}V. Garz\'o, J. Stat. Phys. \textbf{112}, 657--683 (2003).


\bibitem{BT02}A. Barrat and E. Trizac, Granular Matter \textbf{4}, 57--63 (2002).

\bibitem{DHGD02}S. R. Dahl, C. M. Hrenya, V. Garz\'o, and J. W. Dufty, Phys. Rev. E \textbf{66}, 041301 (2002).

\bibitem{puglisi}See for instance, D. Villamaina, A. Puglisi, and A. Vulpiani, J. Stat. Mech. L10001 (2008); A. Sarracino, D. Villamaina, G. Gradenigo, and A. Puglisi, Europhys. Lett. \textbf{92}, 34001 (2010); G. Gradenigo, A. Sarracino, D. Villamaina, and A. Puglisi, J. Stat. Mech. P08017 (2011).


\bibitem{GM02}V. Garz\'o and J. M. Montanero, Physica A \textbf{313}, 336--356 (2002).

\bibitem{GA05}V. Garz\'o and A. Astillero, J. Stat. Phys. \textbf{118}, 935--971 (2005).



\end{thebibliography}


\end{document}